\documentclass{article}
\usepackage{spconf,amsmath,amssymb,graphicx,hyperref}
\usepackage{booktabs}
\usepackage{multirow}
\hypersetup{hidelinks}
\usepackage{xcolor}

\title{TinyAudio: Compact and Efficient Text-to-Audio Generation for Low-Resource Deployment}\name{\shortstack{\itshape Junxi Liu$^{1,2}$, Xiquan Li$^{1,3}$, Wenhao Guan$^{2}$, Yifan Duan$^{1,2}$, Zhikang Niu$^{1,2}$,\\
\itshape Yanru Huo$^{1}$, Ziyang Ma$^{1,2,4}$, Xie Chen$^{1,2,*}$}}

\address{$^{1}$X-LANCE Lab, Shanghai Jiao Tong University, China\\
$^{2}$Shanghai Innovation Institute, China\\
$^{3}$SJTU Paris Elite Institute of Technology, Shanghai Jiao Tong University, China \\
$^{4}$Nanyang Technological University}

\begin{document}
\ninept
\maketitle

\begingroup\renewcommand{\thefootnote}{*}\footnotetext[0]{Corresponding author}\endgroup

\begin{abstract}
Text-to-audio (TTA) generation has advanced rapidly in generation quality and instruction following. However, representative systems often require around a billion parameters, limiting deployment on resource-constrained devices.
This paper introduces \textbf{TinyAudio}, a compact flow-matching-based TTA model for low-resource deployment. At its core, TinyAudio uses TA-DiT, a 35M single-stream flow-matching Transformer. TinyAudio also includes TA-CLAP, a 32M audio-aligned text encoder, and TA-VAE, whose 20M decoder reconstructs 44.1\,kHz audio from compressed latents. 
TinyAudio has only \textbf{87M} parameters in total, over 90\% fewer than representative billion-parameter pipelines, and uses 0.48\,GB peak GPU memory. TinyAudio achieves competitive generation quality on AudioCaps and TTA-Bench. 
\textcolor{black}{We further introduce TinyAudio-MF, a MeanFlow-accelerated model that enables real-time generation with a four-core CPU quota.}
Our results demonstrate a practical quality--footprint trade-off for low-resource TTA deployment.\footnote{Demo: \url{https://tinyaudio-project.github.io/}. Code: \url{https://github.com/junxi25liu/TinyAudio}.}
\end{abstract}


\section{Introduction}
\label{sec:intro}

Text-to-audio (TTA) generation converts natural-language descriptions into realistic acoustic scenes for media production, sound design, and interactive applications. Recent diffusion and flow-matching models \cite{liu2023audioldm,liu2024audioldm2,ghosal2023tango,hung2024tangoflux} have substantially advanced generation quality and text alignment. A typical latent-space TTA system consists of three main components: a text encoder that represents the input description, a generative backbone that produces audio latents conditioned on the text, and a decoder that reconstructs the waveform. High-performing systems often rely on large neural networks and iterative sampling, resulting in substantial memory requirements and inference latency that limit deployment on resource-constrained devices.

\begin{figure}[!t]
    \centering
    \includegraphics[width=\columnwidth]{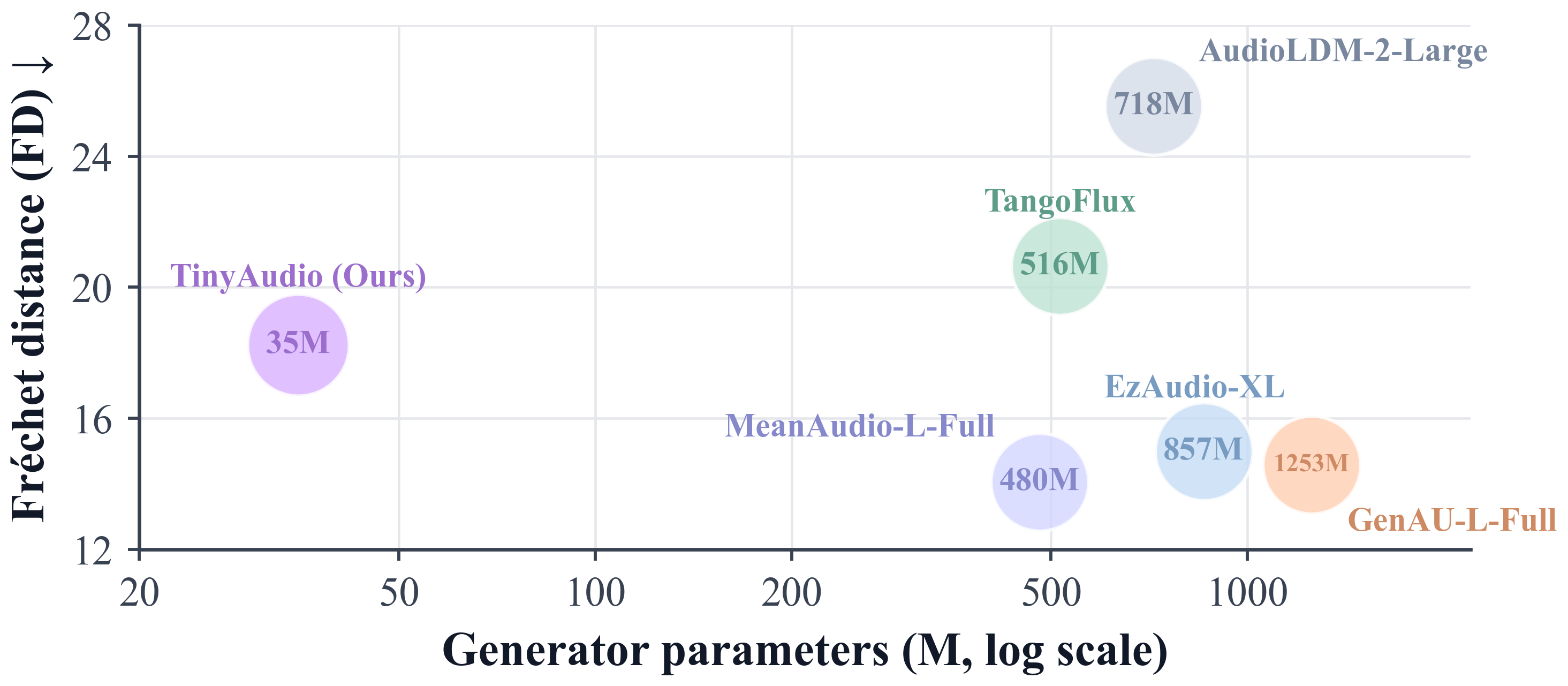}
    \caption{AudioCaps quality versus generator size. Bubble labels indicate generator parameters, and lower FD is better. TinyAudio operates in a substantially smaller parameter regime while retaining competitive distributional quality.}
    \label{fig:quality_footprint}
    \vspace{-0.5cm}
\end{figure}

\begin{figure*}[!t]
    \centering
    \includegraphics[width=0.92\textwidth]{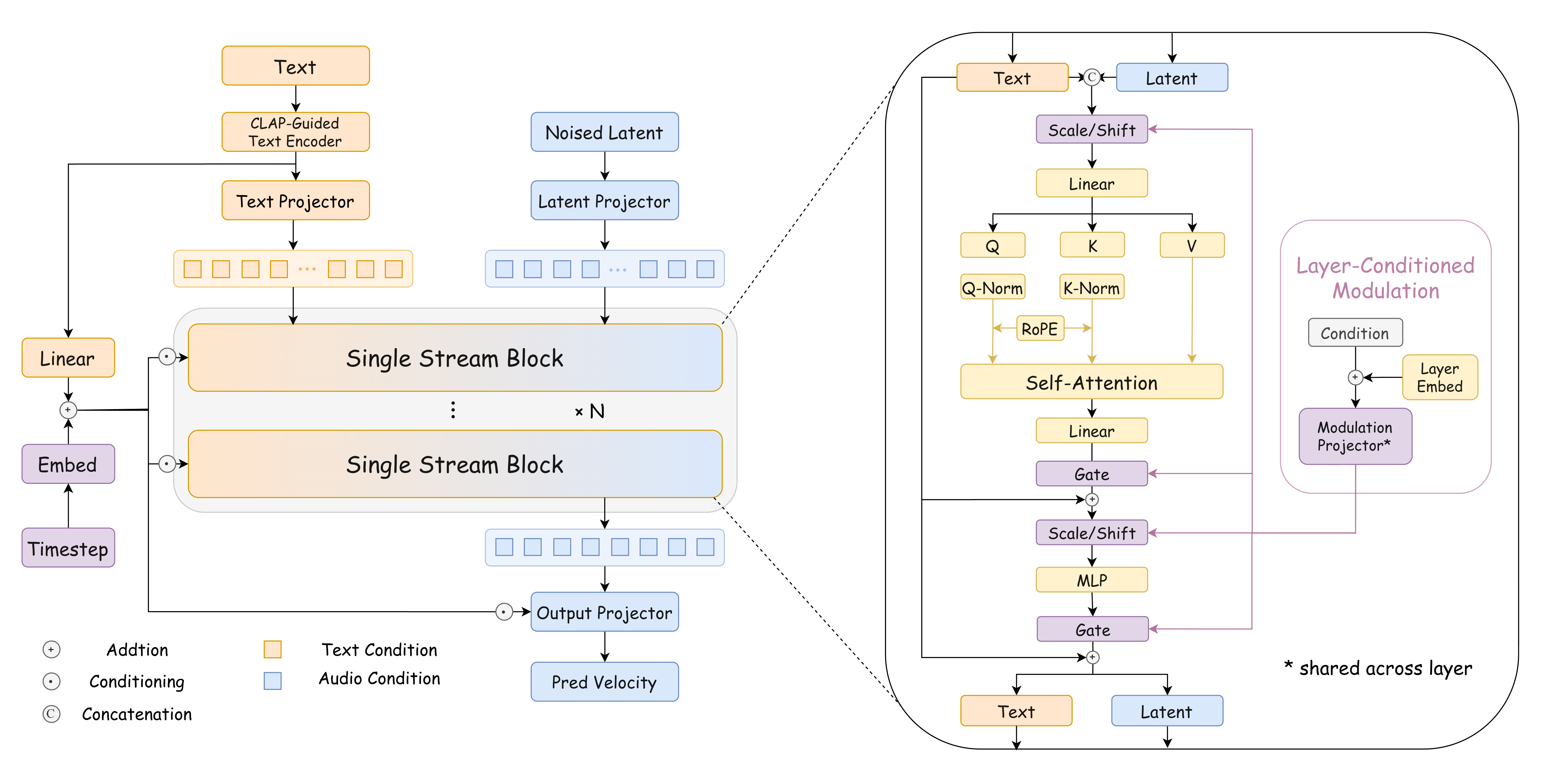}
    \caption{Single-stream TA-DiT. Projected text tokens and noised audio latents are concatenated for joint self-attention. The right panel shows layer-shared conditional modulation: shared MLPs combine text, timestep, and block embeddings to produce shift--scale and gate parameters. Only audio tokens predict latent velocity.}
    \label{fig:ta_dit}
    \vspace{-0.2cm}
\end{figure*}

\begin{table*}[t]
\centering
\small
\setlength{\tabcolsep}{0.5pt}
\begin{tabular*}{\textwidth}{@{\extracolsep{\fill}}lcccccccccccccc@{}}
\toprule
\multirow{2}{*}{\textbf{Model}} &
\multicolumn{5}{c}{\textbf{Footprint / Efficiency}} &
\multicolumn{4}{c}{\textbf{AudioCaps}} &
\multicolumn{5}{c}{\textbf{TTA-Bench}} \\
\cmidrule(lr){2-6}
\cmidrule(lr){7-10}
\cmidrule(lr){11-15}
&
\textbf{Gen.} &
\textbf{Deploy.} &
\textbf{Mem.} &
\textbf{NFE} &
\textbf{RTF} &
\textbf{FAD}$\downarrow$ & \textbf{FD}$\downarrow$ &
\textbf{KL}$\downarrow$ & \textbf{IS}$\uparrow$ &
\textbf{CE}$\uparrow$ & \textbf{CU}$\uparrow$ &
\textbf{PC}$\uparrow$ & \textbf{PQ}$\uparrow$ &
\textbf{CLAP}$\uparrow$ \\
\midrule
AudioLDM-L-Full & 739M & 953M & \underline{3.89\,GB} & 200$\times$2 & 1.190 & 4.32 & 29.50 & 1.68 & 8.17 & 3.275 & 5.137 & 3.219 & 5.854 & 0.441 \\
AudioLDM2-Large & 718M & 1467M & 5.93\,GB & 200$\times$2 & 4.173 & 3.45 & 25.53 & 1.65 & 7.98 & 3.441 & \textbf{5.440} & 2.957 & 5.984 & 0.415 \\
Tango-Full & 866M & 1295M & 5.28\,GB & 200$\times$2 & 1.880 & 2.68 & 15.64 & 1.24 & 8.78 & 3.264 & 5.151 & 3.360 & 5.954 & 0.440 \\
Tango-2-Full & 866M & 1295M & 5.28\,GB & 200$\times$2 & 1.879 & 2.87 & 15.93 & \textbf{1.20} & 10.05 & \underline{3.468} & 5.196 & \textbf{3.800} & 5.895 & 0.467 \\
MeanAudio-L-Full & 480M & 1032M & 4.79\,GB & \underline{25} & \underline{0.057} & 2.59 & \textbf{14.06} & \underline{1.21} & \underline{12.46} & 3.304 & 5.151 & 2.995 & 5.743 & 0.469 \\
EzAudio-XL & 857M & 2120M & 8.61\,GB & 200$\times$2 & 1.896 & 3.64 & 14.98 & 1.29 & 11.38 & 3.388 & 5.055 & 3.663 & 5.719 & 0.446 \\
GenAU-L-Full & 1253M & 2031M & 18.54\,GB & 200$\times$2 & 0.937 & \textbf{2.07} & \underline{14.58} & 1.36 & 10.43 & 3.359 & 5.096 & 3.584 & 5.802 & 0.468 \\
TangoFlux & 516M & \underline{937M} & 6.99\,GB & 50$\times$2 & 0.257 & 2.41 & 20.65 & 1.27 & \textbf{12.81} & \textbf{3.539} & 5.074 & \underline{3.673} & 5.782 & \underline{0.472} \\
Resonate & \underline{470M} & 1108M & 5.13\,GB & 25$\times$2 & 0.273 & \underline{2.27} & 15.53 & 1.31 & 10.79 & 3.451 & \underline{5.328} & 3.232 & \textbf{6.064} & \textbf{0.476} \\
\midrule
\textbf{TinyAudio}
                & \textbf{35M}
                & \textbf{87M}
                & \textbf{0.48\,GB}
                & 25$\times$2
                & 0.078
                & 2.65 & 18.24 & 1.32 & 10.44
                & 3.343 & 5.262 & 2.983 & \underline{5.996} & 0.429 \\
\shortstack{\textbf{TinyAudio-MF}}
                & \textbf{35M}
                & \textbf{87M}
                & \textbf{0.48\,GB}
                & \textbf{1}
                & \textbf{0.010}
                & 2.83 & 23.65 & 1.49 & 8.47
                & 3.350 & 5.257 & 3.043 & 5.994 & 0.403 \\
\bottomrule
\end{tabular*}
\caption{Generation quality, deployment footprint, and inference efficiency. Gen. and Deploy. denote generator and total inference-time parameter counts. Mem. (peak GPU memory during inference) and RTF (end-to-end real-time factor) are measured for all systems on a single GPU in FP32 with batch size 1. NFE counts function evaluations; ${\times}2$ indicates classifier-free guidance. Bold and underline indicate best and second-best results. TTA-Bench results use the Accuracy subset, with baseline values following the Resonate evaluation. \cite{li2026resonate}}
\label{tab:main}
\vspace{-0.2cm}
\end{table*}

Recent approaches such as TangoFlux \cite{hung2024tangoflux} and MeanAudio \cite{li2025meanaudio} improve inference efficiency through faster sampling. However, reducing the number of sampling steps alone does not reduce model size. Building a compact TTA model therefore also requires efficient designs for text conditioning, latent generation, and waveform reconstruction. These components must remain effective as their size decreases: the text encoder needs to capture acoustic semantics, the generator needs to model text--audio relationships, and the decoder needs to reconstruct detailed waveforms from compact latents.

In this paper, we introduce TinyAudio, a compact flow-matching TTA model that addresses these requirements with \textbf{87M} parameters in total. At its core, our 35M TA-DiT jointly processes text and audio tokens in a single stream, avoiding separate modality-specific parameter sets. It further shares conditional modulation networks across Transformer blocks to reduce repeated parameters. For audio-aligned text conditioning, TA-CLAP adapts a 32M text encoder through audio--text contrastive learning \cite{elizalde2023clap,li2026finelap}. For waveform reconstruction, TA-VAE provides a highly compressed audio representation and an approximately 20M decoder that reconstructs 44.1\,kHz waveforms.

Beyond reducing the parameter count, we find that carefully curated training data is critical to the performance of compact TTA models. We therefore introduce a quality-aware SFT strategy that selects audio--text pairs based on perceptual quality and semantic consistency while maintaining balanced sampling across audio domains. To further reduce inference latency, we develop TinyAudio-MF using MeanFlow training, reducing latent generation from 25 Euler steps to a single step.

Experiments on AudioCaps and TTA-Bench demonstrate that TinyAudio achieves competitive generation quality with substantially reduced resource requirements. The standard 25-step TinyAudio model attains an AudioCaps FAD of 2.65 with only 0.48\,GB of peak GPU memory. Its one-step variant, TinyAudio-MF, achieves a steady-state RTF of 0.715 under a four-core CPU quota.

Our contributions are threefold: (1) we introduce \textbf{TinyAudio}, a compact TTA model with only \textbf{87M} inference-time parameters that achieves competitive generation quality; (2) we introduce \textbf{TA-DiT}, which combines single-stream text--audio modeling with layer-shared conditional modulation to improve parameter efficiency; and (3) our ablations show the benefits of parameter-efficient architecture, audio-aligned text conditioning, and quality-aware data selection for compact audio generation.

\section{Method}
\label{sec:method}{\tolerance=1000 \emergencystretch=1em
TinyAudio comprises three core components. First, a 32M-parameter \textcolor{black}{text encoder from TA-CLAP} maps the input prompt to audio-aware representations. Second, a 35M-parameter TA-DiT generates continuous audio latents. Third, a 20M-parameter TA-VAE decoder reconstructs 44.1\,kHz waveforms. Together with lightweight projection layers, TinyAudio contains 87M parameters in total.\par}

\subsection{TA-VAE: Compact Audio Latent Representation}

TinyAudio operates in the compact continuous latent space of \textbf{TA-VAE}. Following SemanticVAE \cite{niu2025semanticvae}, TA-VAE combines a DAC-style convolutional encoder with a lightweight decoder built from anti-aliased multi-periodicity (AMP) blocks. The autoencoder applies a temporal compression factor of $1024\times$ to 44.1\,kHz waveforms, producing a 64-channel latent sequence \textcolor{black}{with 43\,Hz}.

During training, the 7M-parameter encoder produces normalized latent targets; at inference, TA-VAE retains only its approximately 20M-parameter decoder to reconstruct waveforms from generated latents. 

\subsection{TA-CLAP: A Lightweight Audio-Aligned Text Encoder}
\label{subsec:ta_clap}

To strengthen acoustic-semantic representations in a compact 32M text encoder, we introduce \textbf{TA-CLAP}, which adapts the encoder through audio--text contrastive learning. The text encoder is initialized from Ettin-encoder-32m \cite{weller2025seqvsseqopen} and paired with a pretrained HTSAT audio encoder \cite{chen2022htsat} in a dual-encoder architecture \cite{elizalde2023clap}. Lightweight projection heads map the outputs of the audio and text branches into a shared contrastive space.

Given paired audio--text samples, the two branches produce normalized embeddings. We optimize TA-CLAP with the pairwise sigmoid loss introduced in SigLIP \cite{zhai2023sigmoid}.
After contrastive adaptation, the audio branch is discarded. Only the text branch is retained for deployment.\begin{table}[!t]\centering\fontsize{9}{11}\selectfont\setlength{\tabcolsep}{5pt}\begin{tabular}{@{}lcc@{}} \toprule \textbf{Model} & \textbf{OVL}$\uparrow$ & \textbf{REL}$\uparrow$ \\ \midrule \textbf{TinyAudio} & \textbf{3.560{$\pm$0.985}} & \underline{3.520{$\pm$0.940}} \\ TangoFlux & \underline{3.505{$\pm$0.989}} & \textbf{3.600{$\pm$0.992}} \\ EzAudio & 3.155{$\pm$1.169} & 3.190{$\pm$1.169} \\ AudioLDM & 3.015{$\pm$1.004} & 2.730{$\pm$0.962} \\ \bottomrule \end{tabular}\caption{Subjective evaluation of overall quality (OVL) and text relevance (REL). Results are reported as mean $\pm$ standard deviation. Best and second-best scores are shown in bold and underlined.}\label{tab:subjective}
\vspace{-0.3cm}
\end{table}

\subsection{TA-DiT: Lightweight Single-Stream Audio Generation Transformer}

\textbf{TA-DiT} is TinyAudio's compact generative backbone. Its key design objective is to remove two sources of repeated parameters in text-conditioned generative Transformers: separate modality-fusion modules and block-specific conditional modulation networks. TA-DiT addresses them through single-stream audio--text modeling and layer-shared conditional modulation, as illustrated in Figure~\ref{fig:ta_dit}.

The TA-CLAP text encoder produces token-level representations for joint attention and a pooled global condition $c$ for adaptive modulation. TA-DiT first projects the noisy audio latents and text representations into a shared hidden space. Let $A$ and $Y$ denote the projected audio and text token sequences, respectively. They are concatenated along the sequence dimension to form the DiT input:
\begin{equation}
H=[A;Y].
\end{equation}
The joint sequence $H$ is then fed into a stack of Transformer blocks, where audio and text tokens interact through self-attention. In each block, the input sequence is projected into queries, keys, and values. Independently indexed rotary position embeddings \cite{su2021roformer} are applied to the query and key vectors of the audio and text tokens, preserving the positional structure of each modality. This single-stream design uses shared attention parameters across modalities, avoiding separate modality-specific branches and reducing the parameter count of each Transformer block. We additionally apply QK-Norm \cite{henry2020query} to improve training stability.

DiT-style architectures, such as MeanAudio \cite{li2025meanaudio}, typically equip each Transformer block with its own independently parameterized AdaLN modulation network.
This block-specific design introduces substantial parameter overhead. To reduce this overhead, our TA-DiT instead shares conditional modulation networks across all Transformer blocks. Lightweight block-specific embeddings preserve depth-dependent behavior. For block $l$, we combine text and time embeddings as $u=P(c)+E(t)$ and compute
\begin{equation}
    m_{\mathrm{ss}}^{l}=M_{\mathrm{ss}}\!\left(u+e_{\mathrm{ss}}^{l}\right),
\end{equation}
\begin{equation}
    m_{\mathrm{g}}^{l}=M_{\mathrm{g}}\!\left(u+e_{\mathrm{g}}^{l}\right),
\end{equation}
where $P$ projects pooled text features $c$ and $E$ embeds timestep $t$. The subscripts $\mathrm{ss}$ and $\mathrm{g}$ denote shift--scale and gating modulation, respectively. The learned embeddings $e_{\mathrm{ss}}^{l}$ and $e_{\mathrm{g}}^{l}$ are specific to block $l$, while $M_{\mathrm{ss}}$ and $M_{\mathrm{g}}$ are shared across blocks. The outputs $m_{\mathrm{ss}}^{l}$ and $m_{\mathrm{g}}^{l}$ provide adaptive normalization parameters and residual gates for the attention and feed-forward branches, respectively. The attention and feed-forward parameters remain layer-specific.

After the final Transformer block, TA-DiT retains only the audio-token outputs and projects them into latent-velocity predictions for the flow-matching objective \cite{lipman2023flowmatching}.

\subsection{Quality-Aware Supervised Fine-Tuning}
\label{subsec:ta_sft}

We construct the SFT subset by retaining samples with AudioBox Aesthetics scores \cite{tjandra2025audiobox} of PQ $\geq 5.5$, CE $\geq 3.5$, and CU $\geq 4.5$. We group samples into music, speech, and general sounds based on HTSAT predictions \cite{chen2022htsat}, targeting an approximate ratio of 1:1:3. Within each category, we select samples with the highest LAION-CLAP similarity scores \cite{wu2023largescaleclap}. Keeping recordings of 2--30 seconds and deduplicating by audio path yields approximately 417K pairs.

\subsection{MeanFlow One-Step Acceleration}
\label{subsec:meanflow}{\tolerance=1000 \emergencystretch=1em
Standard TinyAudio uses a 25-step Euler solver. Following MeanAudio \cite{li2025meanaudio}, we initialize \textbf{TinyAudio-MF} from the TinyAudio checkpoint pretrained for 500K steps. We then continue training it on the quality-aware SFT subset with the standard SFT optimization settings and the Improved Mean Flows objective \cite{geng2026improvedmeanflows}, reducing sampling to one function evaluation.\par}

\section{Experiments}
\label{sec:experiments}

\subsection{Datasets}

The base training corpus contains approximately 3.7M audio--text pairs from AudioCaps \cite{kim2019audiocaps}, AudioSet \cite{gemmeke2017audioset}, Clotho \cite{drossos2020clotho}, VGGSound \cite{chen2020vggsound}, WavCaps \cite{mei2024wavcaps}, MusicCaps \cite{agostinelli2023musiclm}, and AudioStock\footnote{\url{https://audiostock.net/}}. This corpus supports generator pretraining, TA-CLAP contrastive learning, and TA-VAE reconstruction training. AudioCaps validation and test audio are excluded from both pretraining and SFT. The quality-aware construction pipeline in Section~\ref{subsec:ta_sft} subsequently produces approximately 417K balanced, high-quality pairs for supervised fine-tuning.

We evaluate 10-second generations on the AudioCaps test set \cite{kim2019audiocaps} and evaluate generation quality and audio--text alignment on the Accuracy subset of TTA-Bench \cite{wang2026ttabench}. TTA-Bench contains 1,500 prompts spanning multiple sound events and parallel, sequential, and more complex temporal relations.
\setcounter{topnumber}{1}
\begin{table}[!t]

\centering\small\setlength{\tabcolsep}{2.8pt}\begin{tabular}{@{}p{0.48\columnwidth}ccc@{}} \toprule \textbf{Setting} & \textbf{FAD}$\downarrow$ & \textbf{KL}$\downarrow$ & \textbf{IS}$\uparrow$ \\ \midrule \textbf{Base} & \textbf{2.494} & 1.399 & \textbf{10.695} \\ \midrule \multicolumn{4}{@{}l}{\textsc{Architecture}}\\ Per-block AdaLN & 2.658 & 1.370 & 10.455 \\ Dual-stream MMDiT & 3.168 & \textbf{1.349} & 10.093 \\ \midrule \multicolumn{4}{@{}l}{\textsc{Text Conditioning}}\\ w/o CLAP adaptation & 4.051 & 1.593 & 8.466 \\ \bottomrule \end{tabular}\caption{Component ablations on the AudioCaps test set. All variants are trained for 200K steps on the combined AudioSet and AudioCaps corpus; Base denotes the corresponding complete TA-DiT configuration.}\label{tab:component_ablation_old}

\vspace{2em}
\centering\small\setlength{\tabcolsep}{2.8pt}\begin{tabular}{@{}p{0.48\columnwidth}ccc@{}} \toprule \textbf{Setting} & \textbf{FAD}$\downarrow$ & \textbf{KL}$\downarrow$ & \textbf{IS}$\uparrow$ \\ \midrule Pretraining only & \textbf{2.471} & 1.362 & 10.231 \\ Random & 3.412 & 1.401 & 9.891 \\ \textbf{Quality Aware SFT} & 2.651 & \textbf{1.322} & \textbf{10.442} \\ \bottomrule \end{tabular}\caption{SFT and data-selection ablations on the AudioCaps test set. All variants use the main 3.7M-pair pretraining corpus; the SFT variants use 417K-pair subsets.}\label{tab:sft_ablation}
\vspace{-0.1cm}

\end{table}

\begin{table}[t]

\centering
\small
\setlength{\tabcolsep}{2pt}
\begin{tabular}{@{}lccccccc@{}}
\toprule
Model & Params & \multicolumn{3}{c}{T$\rightarrow$A} & \multicolumn{3}{c}{A$\rightarrow$T} \\
\cmidrule(lr){3-5}\cmidrule(lr){6-8}
 & & R@1 & R@5 & R@10 & R@1 & R@5 & R@10 \\
\midrule
LAION-CLAP & 200M & 35.1 & \textbf{71.9} & \textbf{83.7} & 44.2 & 80.8 & 90.3 \\
TA-CLAP & 67M & \textbf{36.1} & 70.2 & 83.0 & \textbf{48.9} & \textbf{80.9} & \textbf{90.6} \\
\bottomrule
\end{tabular}
\caption{Audio--text retrieval performance of TA-CLAP on AudioCaps test split.}
\label{tab:clap_retrieval}
\vspace{-0.1cm}
\vspace{-0.2cm}
\end{table}

\subsection{Training and Evaluation Details}
TA-CLAP is trained for 10 epochs with a global batch size of 1024, while TA-VAE is trained for 500K steps with a global batch size of 64. TA-DiT uses 18 Transformer blocks with a hidden size of 384 and 8 attention heads. We train the base generator for 500K steps and perform quality-aware SFT for a further 200K steps, using a global batch size of 1024 in both stages. Local and global text conditions are jointly dropped with probability 0.1 to learn the unconditional field used for classifier-free guidance. Standard TinyAudio uses a guidance strength of 9 and a 25-step Euler solver.

Both stages use fused AdamW with peak learning rates of $1\times10^{-4}$ and $5\times10^{-5}$, weight decay $1\times10^{-6}$, 1{,}000-step warmup, gradient clipping at 1.0, and MultiStepLR milestones $[360\text{K},430\text{K}]$ and $[160\text{K},180\text{K}]$ for pretraining and SFT, respectively, with the learning rate multiplied by 0.1 at each milestone.

On AudioCaps, we report Fr\'echet Audio Distance (FAD), Fr\'echet Distance (FD), KL divergence, and Inception Score (IS). FAD uses VGGish features \cite{kilgour2019frechet}, where FD, KL, and IS use PANNs features \cite{kong2020panns}. On TTA-Bench, we report the AudioBox-Aesthetics dimensions Content Enjoyment (CE), Content Usefulness (CU), Production Complexity (PC), and Production Quality (PQ), together with Resonate-style CLAP similarity \cite{li2026resonate}.

For subjective evaluation, ten audio experts each rate 10 samples per model on five-point scales for overall quality (OVL) and text relevance (REL). Samples are presented in random order with model identities hidden. All participants provide informed consent.\subsection{Generation Quality and Deployment Cost}

\textbf{Generation quality.} Table~\ref{tab:main} compares TinyAudio with representative TTA systems~\cite{majumder2024tango2,hai2025ezaudio,haji2026taming}. TinyAudio achieves an AudioCaps FAD of 2.65, close to Tango and MeanAudio, with competitive KL but higher FD than most compared systems. On TTA-Bench, TinyAudio is competitive in CU and PQ. TinyAudio-MF substantially reduces sampling cost, but its objective generation metrics indicate a measurable quality trade-off. In Table~\ref{tab:subjective}, TinyAudio ranks first in mean OVL and second in mean REL, behind TangoFlux.

\noindent\textbf{Deployment cost.} TinyAudio uses 0.48\,GB of peak GPU memory, more than $8\times$ less than every compared system. CPU inference is evaluated on a cluster node equipped with two Intel Xeon Gold 6530 processors, with the container limited via cgroup to a CPU quota equivalent to four cores. TinyAudio-MF achieves a mean steady-state generation RTF of 0.715 for 10-second audio in FP32 with batch size 1, averaged over 5 runs after 10 warm-up runs. Timing includes text encoding and waveform decoding but excludes model loading and audio saving.

\subsection{Ablation Studies}

\noindent\textbf{Architecture and text conditioning.} Table~\ref{tab:component_ablation_old} compares architectural and text-conditioning variants trained for 200K steps on AudioSet and AudioCaps. Per-block AdaLN uses block-specific modulation, while Dual-stream adopts MMDiT. Both variants use fewer Transformer blocks to match Base's parameter count. All models are evaluated on 957 AudioCaps test samples with 25-step Euler sampling and a guidance strength of 9. At matched parameter counts, Base offers a better overall quality trade-off, supporting single-stream modeling and shared modulation as parameter-efficient choices for compact generators. Removing TA-CLAP adaptation consistently degrades performance, showing the importance of audio-aligned text representations when generator capacity is limited.

\noindent\textbf{SFT and data selection.} Table~\ref{tab:sft_ablation} compares pretraining only, Random SFT, and SFT under the main training setup. Random SFT uses 417K pairs randomly sampled from the pretraining corpus. Both SFT variants use equally sized subsets with identical initialization and optimization settings. For compact generators, additional fine-tuning alone does not guarantee improvement. Quality-aware data selection is more effective than random sampling.

\noindent\textbf{TA-CLAP's performance.} Table~\ref{tab:clap_retrieval} reports audio--text retrieval performance (\%) on the AudioCaps test split, with LAION-CLAP results re-evaluated by us. Parameter counts include both audio and text branches. 
Overall, TA-CLAP achieves comparable retrieval performance to LAION-CLAP \cite{wu2023largescaleclap} in both directions, demonstrating strong alignment between audio and text modalities.

\section{Conclusion}

We introduced TinyAudio, a compact flow-matching TTA model with 87M inference-time parameters. Its core TA-DiT combines single-stream text--audio modeling with layer-shared conditional modulation. TA-CLAP provides audio-aligned text conditioning, while TA-VAE reconstructs 44.1\,kHz waveforms from compressed latents. TinyAudio achieves competitive generation quality on AudioCaps and TTA-Bench with 0.48\,GB peak GPU memory. TinyAudio-MF further enables real-time generation with a four-core CPU quota. These results demonstrate a practical quality--footprint trade-off for low-resource TTA deployment.\clearpage
\section{Compliance with Ethical Standards}
The subjective listening test was conducted with the informed consent of all
participants, who were compensated for their time. This study was conducted
in accordance with the principles of the Declaration of Helsinki.

\bibliographystyle{IEEEbib}
\bibliography{tinyaudio}

@article{liu2023audioldm,
  title={{AudioLDM}: Text-to-audio generation with latent diffusion models},
  author={Liu, Haohe and Chen, Zehua and Yuan, Yi and Mei, Xinhao and Liu, Xubo and Mandic, Danilo and Wang, Wenwu and Plumbley, Mark D},
  journal={arXiv preprint arXiv:2301.12503},
  year={2023}
}

@article{liu2024audioldm2,
  title={{AudioLDM} 2: Learning holistic audio generation with self-supervised pretraining},
  author={Liu, Haohe and Yuan, Yi and Liu, Xubo and Mei, Xinhao and Kong, Qiuqiang and Tian, Qiao and Wang, Yuping and Wang, Wenwu and Wang, Yuxuan and Plumbley, Mark D},
  journal={IEEE/ACM Trans. Audio, Speech, Lang. Process.},
  volume={32},
  pages={2871--2883},
  year={2024}
}

@inproceedings{ghosal2023tango,
  title={Text-to-audio generation using instruction guided latent diffusion model},
  author={Ghosal, Deepanway and Majumder, Navonil and Mehrish, Ambuj and Poria, Soujanya},
  booktitle={Proc. ACM MM},
  pages={3590--3598},
  year={2023}
}

@inproceedings{hung2024tangoflux,
  title={{TangoFlux}: Super fast and faithful text to audio generation with flow matching and clap-ranked preference optimization},
  author={Hung, Chia-Yu and Majumder, Navonil and Kong, Zhifeng and Mehrish, Ambuj and Zadeh, Amir and Li, Chuan and Valle, Rafael and Catanzaro, Bryan and Poria, Soujanya},
  booktitle={Proc. ICLR},
  year={2026}
}

@inproceedings{li2025meanaudio,
  title={{MeanAudio}: Fast and faithful text-to-audio generation with mean flows},
  author={Li, Xiquan and Liu, Junxi and Liang, Yuzhe and Niu, Zhikang and Chen, Wenxi and Chen, Xie},
  booktitle={Proc. ACL},
  pages={14378--14393},
  year={2026}
}

@inproceedings{elizalde2023clap,
  title={{CLAP}: Learning audio concepts from natural language supervision},
  author={Elizalde, Benjamin and Deshmukh, Soham and Al Ismail, Mahmoud and Wang, Huaming},
  booktitle={Proc. ICASSP},
  pages={1--5},
  year={2023}
}

@inproceedings{wu2023largescaleclap,
  title={Large-scale contrastive language-audio pretraining with feature fusion and keyword-to-caption augmentation},
  author={Wu, Yusong and Chen, Ke and Zhang, Tianyu and Hui, Yuchen and Berg-Kirkpatrick, Taylor and Dubnov, Shlomo},
  booktitle={Proc. ICASSP},
  pages={1--5},
  year={2023}
}

@article{niu2025semanticvae,
  title={{Semantic-VAE}: Semantic-alignment latent representation for better speech synthesis},
  author={Niu, Zhikang and Hu, Shujie and Choi, Jeongsoo and Chen, Yushen and Chen, Peining and Zhu, Pengcheng and Yang, Yunting and Zhang, Bowen and Zhao, Jian and Wang, Chunhui and others},
  journal={arXiv preprint arXiv:2509.22167},
  year={2025}
}

@inproceedings{weller2025seqvsseqopen,
  title={Seq vs seq: An open suite of paired encoders and decoders},
  author={Weller, Orion and Ricci, Kathryn and Marone, Marc and Chaffin, Antoine and Lawrie, Dawn and Van Durme, Benjamin},
  booktitle={Proc. ICLR},
  year={2026}
}

@inproceedings{chen2022htsat,
  title={{HTS-AT}: A hierarchical token-semantic audio transformer for sound classification and detection},
  author={Chen, Ke and Du, Xingjian and Zhu, Bilei and Ma, Zejun and Berg-Kirkpatrick, Taylor and Dubnov, Shlomo},
  booktitle={Proc. ICASSP},
  pages={646--650},
  year={2022}
}

@inproceedings{zhai2023sigmoid,
  title={Sigmoid loss for language image pre-training},
  author={Zhai, Xiaohua and Mustafa, Basil and Kolesnikov, Alexander and Beyer, Lucas},
  booktitle={Proc. ICCV},
  pages={11941--11952},
  year={2023}
}

@article{su2021roformer,
  title={{RoFormer}: Enhanced transformer with rotary position embedding},
  author={Su, Jianlin and Ahmed, Murtadha and Lu, Yu and Pan, Shengfeng and Bo, Wen and Liu, Yunfeng},
  journal={Neurocomputing},
  volume={568},
  pages={127063},
  year={2024}
}

@inproceedings{henry2020query,
  title={Query-key normalization for transformers},
  author={Henry, Alex and Dachapally, Prudhvi Raj and Pawar, Shubham Shantaram and Chen, Yuxuan},
  booktitle={Findings of EMNLP},
  pages={4246--4253},
  year={2020}
}

@article{lipman2023flowmatching,
  title={Flow matching for generative modeling},
  author={Lipman, Yaron and Chen, Ricky T. Q. and Ben-Hamu, Heli and Nickel, Maximilian and Le, Matt},
  journal={arXiv preprint arXiv:2210.02747},
  year={2022}
}

@article{tjandra2025audiobox,
  title={Meta {AudioBox} Aesthetics: Unified automatic quality assessment for speech, music, and sound},
  author={Tjandra, Andros and Wu, Yi-Chiao and Guo, Baishan and Hoffman, John and Ellis, Brian and Vyas, Apoorv and Shi, Bowen and Chen, Sanyuan and Le, Matt and Zacharov, Nick and others},
  journal={arXiv preprint arXiv:2502.05139},
  year={2025}
}

@inproceedings{geng2026improvedmeanflows,
  title={Improved mean flows: On the challenges of fastforward generative models},
  author={Geng, Zhengyang and Lu, Yiyang and Wu, Zongze and Shechtman, Eli and Kolter, J. Zico and He, Kaiming},
  booktitle={Proc. CVPR},
  pages={30467--30476},
  year={2026}
}

@inproceedings{kim2019audiocaps,
  title={{AudioCaps}: Generating captions for audios in the wild},
  author={Kim, Chris Dongjoo and Kim, Byeongchang and Lee, Hyunmin and Kim, Gunhee},
  booktitle={Proc. NAACL},
  pages={119--132},
  year={2019}
}

@inproceedings{gemmeke2017audioset,
  title={{AudioSet}: An ontology and human-labeled dataset for audio events},
  author={Gemmeke, Jort F. and Ellis, Daniel P. W. and Freedman, Dylan and Jansen, Aren and Lawrence, Wade and Moore, R. Channing and Plakal, Manoj and Ritter, Marvin},
  booktitle={Proc. ICASSP},
  pages={776--780},
  year={2017}
}

@inproceedings{drossos2020clotho,
  title={Clotho: An audio captioning dataset},
  author={Drossos, Konstantinos and Lipping, Samuel and Virtanen, Tuomas},
  booktitle={Proc. ICASSP},
  pages={736--740},
  year={2020}
}

@inproceedings{chen2020vggsound,
  title={{VGGSound}: A large-scale audio-visual dataset},
  author={Chen, Honglie and Xie, Weidi and Vedaldi, Andrea and Zisserman, Andrew},
  booktitle={Proc. ICASSP},
  pages={721--725},
  year={2020}
}

@article{mei2024wavcaps,
  title={{WavCaps}: A ChatGPT-assisted weakly-labelled audio captioning dataset for audio-language multimodal research},
  author={Mei, Xinhao and Meng, Chutong and Liu, Haohe and Kong, Qiuqiang and Ko, Tom and Zhao, Chengqi and Plumbley, Mark D. and Zou, Yuexian and Wang, Wenwu},
  journal={IEEE/ACM Trans. Audio, Speech, Lang. Process.},
  volume={32},
  pages={3339--3354},
  year={2024}}

@inproceedings{majumder2024tango2,
  title={{Tango 2}: Aligning diffusion-based text-to-audio generations through direct preference optimization},
  author={Majumder, Navonil and Hung, Chia-Yu and Ghosal, Deepanway and Hsu, Wei-Ning and Mihalcea, Rada and Poria, Soujanya},
  booktitle={Proc. ACM MM},
  pages={564--572},
  year={2024}
}

@inproceedings{hai2025ezaudio,
  title={{EzAudio}: Enhancing text-to-audio generation with efficient diffusion transformer},
  author={Hai, Jiarui and Xu, Yong and Zhang, Hao and Li, Chenxing and Wang, Helin and Elhilali, Mounya and Yu, Dong},
  booktitle={Proc. Interspeech},
  pages={4233--4237},
  year={2025}
}

@article{haji2026taming,
  title={Taming data and transformers for audio generation},
  author={Haji-Ali, Moayed and Menapace, Willi and Siarohin, Aliaksandr and Balakrishnan, Guha and Ordonez, Vicente},
  journal={Int. J. Comput. Vis.},
  volume={134},
  number={3},
  pages={87},
  year={2026}
}

@inproceedings{li2026finelap,
  title={Finelap: Taming heterogeneous supervision for fine-grained language-audio pretraining},
  author={Li, Xiquan and Xu, Xuenan and Ma, Ziyang and Chen, Wenxi and He, Haolin and Kong, Qiuqiang and Chen, Xie},
  booktitle={Proc. ACL},
  year={2026}
}

@article{agostinelli2023musiclm,
  title={{MusicLM}: Generating music from text},
  author={Agostinelli, Andrea and Denk, Timo I. and Borsos, Zal{\'a}n and Engel, Jesse and Verzetti, Mauro and Caillon, Antoine and Huang, Qingqing and Jansen, Aren and Roberts, Adam and Tagliasacchi, Marco and others},
  journal={arXiv preprint arXiv:2301.11325},
  year={2023}
}

@inproceedings{wang2026ttabench,
  title={{TTA-Bench}: A comprehensive benchmark for evaluating text-to-audio models},
  author={Wang, Hui and Liu, Cheng and Chen, Junyang and Liu, Haoze and Jia, Yuhang and Zhao, Shiwan and Zhou, Jiaming and Sun, Haoqin and Bu, Hui and Qin, Yong},
  booktitle={Proc. AAAI},
  year={2026}
}

@article{kilgour2019frechet,
  title={Fr{\'e}chet audio distance: A metric for evaluating music enhancement algorithms},
  author={Kilgour, Kevin and Zuluaga, Mauricio and Roblek, Dominik and Sharifi, Matthew},
  journal={arXiv preprint arXiv:1812.08466},
  year={2018}
}

@article{kong2020panns,
  title={{PANNs}: Large-scale pretrained audio neural networks for audio pattern recognition},
  author={Kong, Qiuqiang and Cao, Yin and Iqbal, Turab and Wang, Yuxuan and Wang, Wenwu and Plumbley, Mark D.},
  journal={IEEE/ACM Trans. Audio, Speech, Lang. Process.},
  volume={28},
  pages={2880--2894},
  year={2020}
}

@inproceedings{li2026resonate,
  title={Resonate: Reinforcing text-to-audio generation via online feedback from large audio language models},
  author={Li, Xiquan and Liu, Junxi and Chen, Wenxi and Zhu, Haina and Ma, Ziyang and Chen, Xie},
  booktitle={Proc. Interspeech},
  year={2026}
}

\end{document}